\documentclass[aps,prl,nofootinbib,showpacs,twocolumn,superscriptaddress]{revtex4-2}

\usepackage[english]{babel}
\usepackage[utf8]{inputenc}
\usepackage{amsmath}
\usepackage{amssymb}
\usepackage[caption=false]{subfig}
\usepackage{amssymb}
\usepackage{epsfig}
\usepackage{graphicx}
\usepackage{amsmath}
\usepackage{array,color}
\usepackage{natbib}

\usepackage[usenames,dvipsnames]{xcolor}
\definecolor{forestgreen}{rgb}{0.11,0.54,0.15}
\definecolor{purple}{rgb}{0.62,0.10,0.96}
\definecolor{dockerblue}{rgb}{0.11,0.56,0.98}
\definecolor{freeblue}{rgb}{0.25,0.41,0.88}

\usepackage[pdftex,plainpages=false,colorlinks=true,linkcolor=Red, citecolor=blue, urlcolor=blue]{hyperref}
\begin{document}

\makeatletter
\def\fnum@figure{Fig.~\thefigure}
\makeatother

\title{Quantization through dissipation: impurity states and the optical quantum Hall effect}

\author{Zhenisbek Tagay}
\affiliation{Department of Physics and Astronomy, The Johns Hopkins University, Baltimore, MD 21218 USA.}

\author{Ahmed Abouelkomsan}
\affiliation{Department of Physics, Massachusetts Institute of Technology, Cambridge, Massachusetts 02139, USA}

\author{Yugo Onishi}
\affiliation{Department of Physics, Massachusetts Institute of Technology, Cambridge, Massachusetts 02139, USA}

\author{Adbhut Gupta}
\affiliation{Department of Electrical and Computer Engineering, Princeton University, Princeton, NJ 08854 USA.}

\author{Loren Pfeiffer}
\affiliation{Department of Electrical and Computer Engineering, Princeton University, Princeton, NJ 08854 USA.}

\author{Liang Fu}
\affiliation{Department of Physics, Massachusetts Institute of Technology, Cambridge, Massachusetts 02139, USA}
\affiliation{Canadian Institute for Advanced Research, Toronto, Ontario, Canada}

\author{N.~P.~Armitage}
\email{npa@jhu.edu}
\affiliation{Department of Physics and Astronomy, The Johns Hopkins University, Baltimore, MD 21218 USA.}
\affiliation{Canadian Institute for Advanced Research, Toronto, Ontario, Canada}

\date{\today}

\maketitle

\begin{figure*}[t!]
\centering
\includegraphics[width=0.8\textwidth]{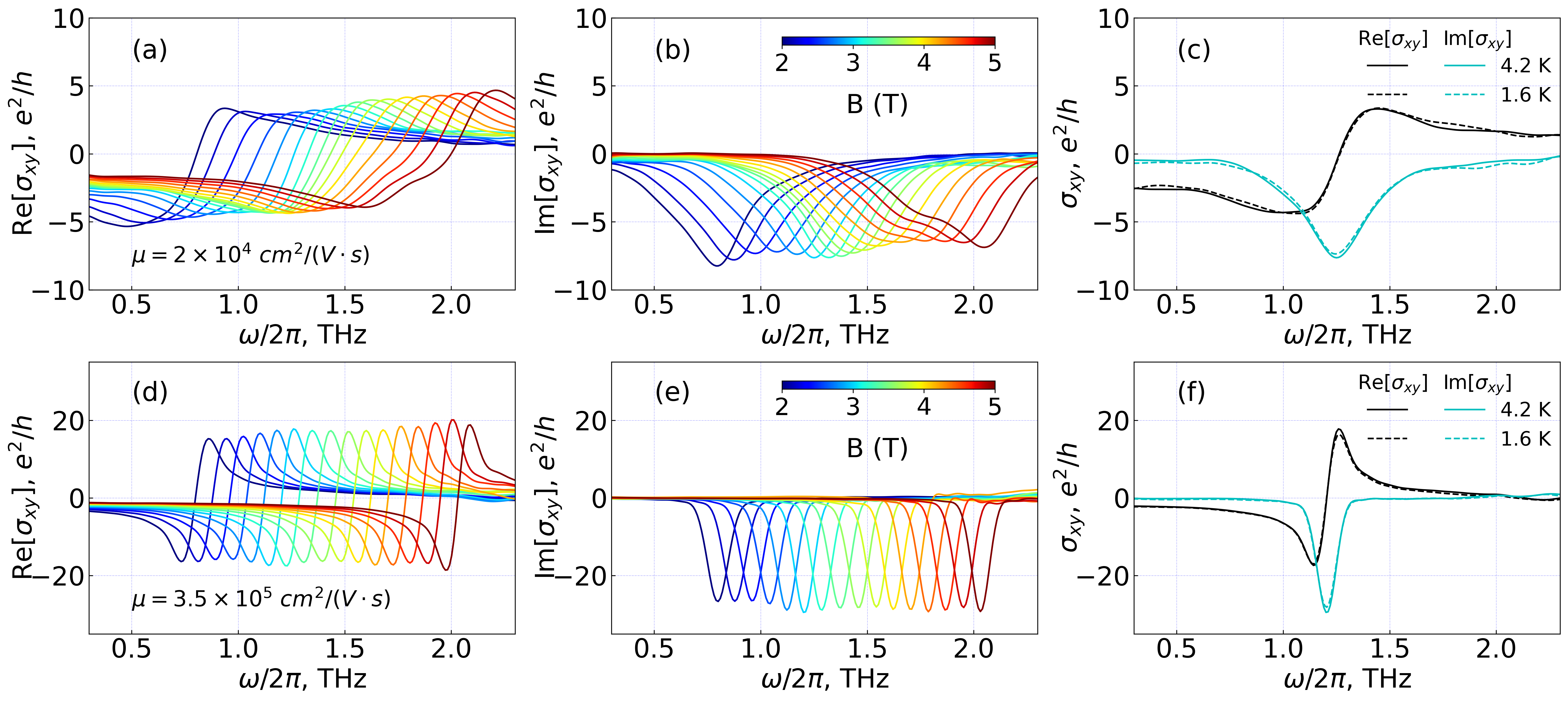} 
\label{Fig.Conductivity}
\caption{{\bf THz range complex Hall conductivity for 2DEG} (a) Real and (b) imaginary Hall conductance $\sigma_{xy}(\omega)$ of lower-mobility 2DEG at T~=~4.2~K and various magnetic fields. (c) Comparison between $\sigma_{xy}(\omega)$ measured at T~=~4.2~K and T~=~1.6~K. (d)-(f) Same plots for higher-mobility 2DEG.}
\end{figure*}

{\bf The integer quantum Hall effect remains the most paradigmatic and remarkable example of exact quantization in condensed matter physics~\cite{klitzing1980new,girvin1999quantum}. Its Hall conductivity $\sigma_{xy}$ is fixed to $e^2/h$ times an integer to a precision limited only by measurement and independent of disorder or interactions. This robustness has several explanations, each capturing a strikingly different piece of the physics. In Laughlin's gauge argument~\cite{Laughlin}, flux insertion pumps an integer charge between edges, so quantization follows from gauge invariance alone. The edge-channel picture~\cite{Halperin,Buttiker} instead attributes transport to chiral, ballistic 1D channels at the sample boundary, one per filled Landau level. Bulk arguments tie $\sigma_{xy}$ to a topological invariant of the disordered system~\cite{TKNN,NiuThoulessWu}, with extended states compensating exactly for the current not carried by localized ones~\cite{Prange,AokiAndo}. Here we demonstrate an additional route to understanding quantization, rooted in the finite-frequency dissipative electrodynamics of the bulk. Using high-precision terahertz Faraday rotation and numerics, we show that the cyclotron resonance alone does not give quantized plateaus under Kramers-Kronig transformation. Quantization is recovered only once a faint, low-frequency, topologically enforced dissipative contribution from impurity states is included. This dynamical mechanism, hiding in plain sight within the dissipative response, ties the DC value to the finite-frequency optical response and offers a new route to quantization through the optical Hall effect, complementary to previous pictures.}

In these experiments we measured the THz range Hall conductivity through high precision THz range Faraday rotation measurements. The dissipative spectrum is dominated by the classical cyclotron resonance, a finite frequency peak that moves smoothly as a function of field as $\omega_c = eB/m$. In principle, one can always obtain the DC Hall response via a Kramers-Kronig (KK) integral transform of the dissipative response. However, this presents a bit of a puzzle as the cyclotron resonance spectra does not have enough structure to reproduce the quantized DC response. We show that the plateau regions show up in the cyclotron range spectra only as Shubnikov-de Haas like oscillations in the Drude transport parameters in low mobility samples (and not at all in the higher mobility ones).   The KK transform of this data produces only subtle rounded step features, but no quantized plateaus.  We show numerically that quantization is recovered only when a low-frequency dissipative contribution from impurity states is included in the Kramers-Kronig integral. This contribution is topologically enforced, supplying precisely the correction needed to reproduce the quantized Hall conductance.

GaAs/AlGaAs-based 2D electron gases (2DEGs) have long served as a premier platform for the observation of the integer \cite{AlGaAs_QHE} and, subsequently, fractional \cite{AlGaAs_FQHE} quantum Hall effects in DC transport measurements. Signatures of the quantum Hall-like effect have also been observed in microwave \cite{Volkov, Schlapp_Microwave2, Shlapp_Microwave3, Schlapp_Microwave4, Bertrand_microwave, Arakawa_Microwave} and THz \cite{Shimano_AlGaAs, Pimenov_AlGaAs} measurements, although their manifestation at finite frequencies is more nuanced. To probe the low-energy quantum Hall regime, the probing frequency should remain well below the cyclotron frequency, $\omega\ll\omega_c$, as the Landau-level spacing, $\hbar\omega_c$, sets the relevant energy scale. At the same time, finite frequency introduces an additional energy scale that, similarly to temperature, broadens the transition between quantum Hall plateaus through dynamical scaling effects \cite{Hohls_scaling, engel_scaling}. Experimentally, accessing this regime is further complicated by the difficulty of combining high-frequency measurements with very low temperatures.

To investigate this topic, we measure GaAs/AlGaAs two-dimensional electron gas (2DEG) systems using DC and THz-frequency techniques. Experimental details are provided in Methods section. The 2DEGs were grown by molecular beam epitaxy on $0.5$-mm-thick GaAs (001) substrates and consist of single-sided modulation-doped GaAs quantum wells (QWs) with Al$_{0.32}$Ga$_{0.68}$As barriers. We study two samples with different levels of disorder, having mobilities of $3.5 \times 10^{5}~\mathrm{cm^2/(V\cdot s)}$ and $2 \times 10^{4}~\mathrm{cm^2/(V\cdot s)}$ measured at $0.3~\mathrm{K}$ shortly after the synthesis. Under the conditions of the present measurements, at $T=1.5~\mathrm{K}$, the carrier densities were $1.4\times10^{11}~\mathrm{cm^{-2}}$ and $1.65\times10^{11}~\mathrm{cm^{-2}}$ for the higher- and lower-mobility samples, respectively. 
In the higher-mobility sample, the $40$-nm-wide QW is separated from the Si $\delta$-doped layer by a $70$-nm spacer, with an additional Si $\delta$-layer with a doping density of $\sim 3 \times 10^{11}~\mathrm{cm^{-2}}$ placed $10~\mathrm{nm}$ from the QW to enhance remote-ionized impurity scattering. The lower-mobility sample contains a $30$-nm-wide QW with the Si $\delta$-doped layer $54~\mathrm{nm}$ away, while Si is also incorporated directly into the QW at a concentration of $\sim 2 \times 10^{16}~\mathrm{cm^{-3}}$ to introduce additional short-range disorder.

In Figs.~1(a)-(b), we show the real and imaginary parts of $\sigma_{xy}(\omega)$ measured at $T=4.2$ K and magnetic fields between 2 and 5 T for a 2DEG sample with a lower carrier mobility of $\mu=2\times10^4~\mathrm{cm^2/(V\cdot s)}$. The frequency and field dependence of both Re[$\sigma_{xy}(\omega)$] and Im[$\sigma_{xy}(\omega)$] are well described by the semi-classical Drude model with cyclotron resonance (CR), with the absorption peak shifting linearly with the applied magnetic field ($\omega_c=eB/m$). The corresponding scattering rate of this resonance $\Gamma$ (half-width at half maximum) is estimated to be $\approx0.2$ THz. At higher magnetic fields, an additional absorption feature emerges at $\omega<\omega_c$, becoming clearly resolved above approximately $4$ T. With increasing magnetic field, the separation between this feature and the main cyclotron resonance gradually increases. Similar additional resonances in the vicinity of the cyclotron resonance have previously been observed in GaAs/AlGaAs 2DEGs and associated with impurity-mediated collective excitations \cite{Schlesinger, Henriksen} as well as magneto-optical transitions involving impurity-bound states~\cite{Chang, Huant}.

In Figs.~1(d)-(e), we show $\sigma_{xy}(\omega)$ for the higher-mobility sample with $\mu=3.5\times10^5~\mathrm{cm^2/(V\cdot s)}$. The data are also well described by the semi-classical model, but with a much sharper cyclotron resonance ($\Gamma\approx0.05$ THz) reflecting the higher mobility. In contrast to the lower-mobility sample, no additional absorption is observed at $\omega<\omega_c$. The presence of this feature only in the lower-mobility sample further suggests an important role of impurities in its origin. In Figs.~1(c) and 1(f), we compare the data measured at $T=1.6$ K and $T=4.2$ K for both samples. Neither the real nor the imaginary part shows any significant difference between the two temperatures at $B=3$ T. Full magnetic dependence of $\sigma_{xy}(\omega)$ at $T=1.6$ K is provided in Supplementary Fig.~S2. 

The smoothly evolving spectra of Fig. 1 contrast with the notable quantized plateaus found around 3.5 Tesla in the DC response (Fig. 3) for both samples. Even at the lowest end of our spectral range (see $0.3$ THz in Fig. 3) we do not observe plateaus. It is only at higher frequencies closer to the cyclotron resonance that a plateau-like behavior in Re[$\sigma_{xy}$] emerges for the lower-mobility sample. In Fig.~2(a), we show a $\omega/2\pi=0.6$ THz frequency cut of Re[$\sigma_{xy}(\omega)$] at $T=1.6$ K and $T=4.2$ K as a function of the applied magnetic field $B$. In contrast to the expected semi-classical $1/B$ dependence, at both temperatures we observe a plateau at $\approx2.5~e^2/h$ between 3 and 4 T. The center of this feature coincides with that of the $\nu=2$ plateau observed in DC Hall transport (see Fig.~3). At $T=1.6$ K, the plateau is slightly wider than at $T=4.2$ K, although in both cases its width is approximately the same as in the corresponding DC transport data.
\begin{figure}[t!]
\centering
\includegraphics[width=0.48\textwidth]{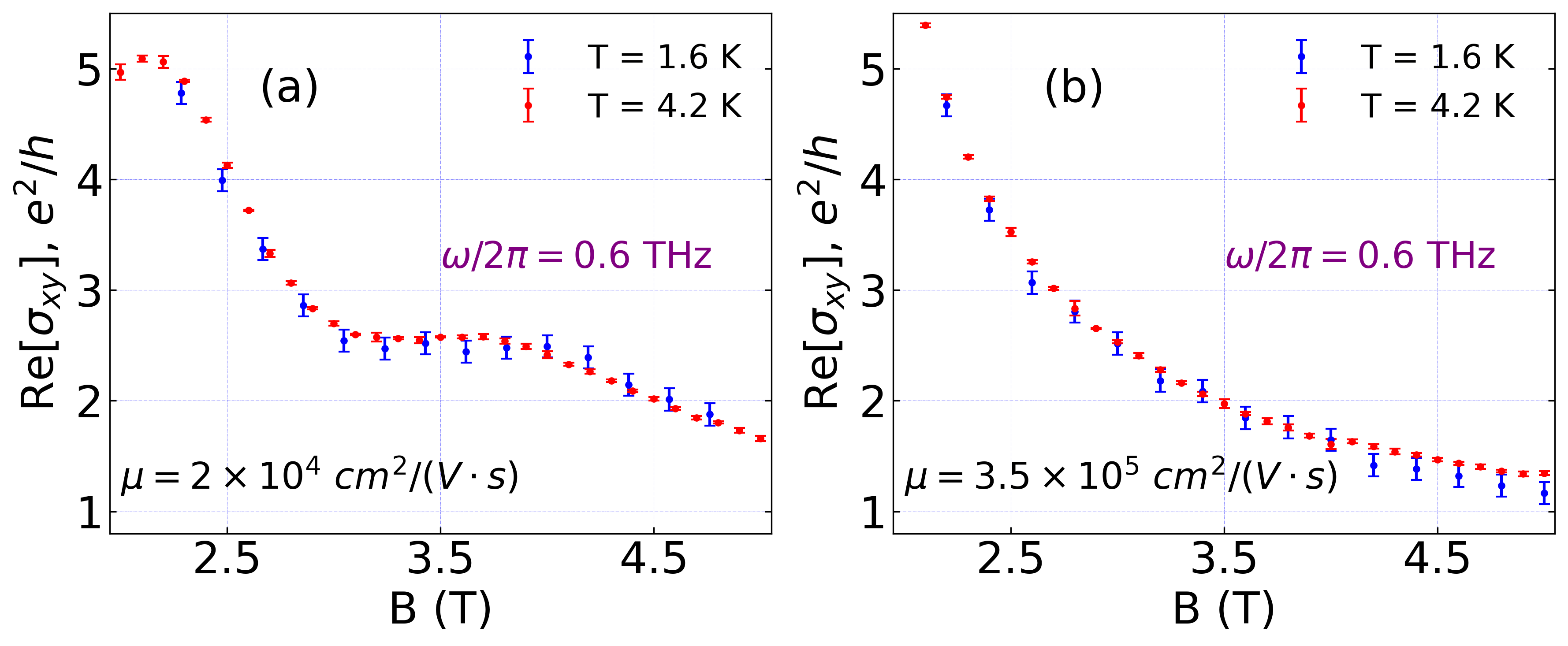}
\label{Fig.Conductivity}
\caption{{\bf Magnetic field dependence of Re[$\sigma_{xy}$] at $\omega/2\pi$~=~0.6~THz} for (a) lower-mobility and (b) higher-mobility samples at two different temperatures.}
\end{figure}
Previous theoretical studies have shown that non-quantized plateau-like features can persist at finite frequencies due to disorder-induced localization, with their values deviating from the DC quantized value~\cite{Morimoto_2009, Morimoto_2010}. In the cyclotron resonance regime, Landau-level filling can additionally manifest itself through Shubnikov--de Haas-like oscillations of the cyclotron resonance parameters, including its effective mass and linewidth~\cite{Englert,Richter,Besson,Kono}. Such oscillations can be sensitive to impurity scattering~\cite{Richter}, and the plateau-like behavior observed here may reflect their interplay with the finite-frequency quantum Hall response. Regardless of its microscopic origin, it is notable that this feature appears at the same magnetic field as the DC quantum Hall plateau and extends over approximately the same field range. Additional frequency cuts showing the evolution of this feature are presented in the Supplementary Figs.~S3-S6.
In contrast, the higher-mobility sample follows the semi-classical $1/B$ dependence closely, as shown in Fig.~2(b), and does not show any clear plateau-like features at these frequencies.

As mentioned above, we do not observe plateau-like behavior in Re[$\sigma_{xy}(\omega)$] at the lowest end of our spectral range. To compare these results directly with the DC response, in Fig.~3 we plot the Hall conductance $\sigma_{xy}^{DC}$ (solid black line), measured simultaneously with the THz experiment, as well as the simulated semi-classical $\sigma_{xy}^{\mathrm{class}}$ curve (dashed black line) based on the fitted Drude parameters. For both samples, Re[$\sigma_{xy}(0.3 \; \mathrm{THz})$] not only lacks plateau-like signatures but is even larger than the corresponding DC value (as expected within the semi-classical picture, since the finite-frequency response acquires an $\omega^2$ correction to its DC value~\cite{Ethan}).

Since $0.3$ THz is the lower measured end of our spectral range, the quantum Hall response clearly observed in DC transport must emerge at even lower frequencies. Importantly, through the Kramers-Kronig relation, any additional contribution to Re[$\sigma_{xy}(\omega)$] required to recover the DC quantum Hall plateaus must have a corresponding signature in Im[$\sigma_{xy}(\omega)$] as the zero-frequency Hall conductivity is related to the imaginary part by the KK relation

\begin{equation}
\mathrm{Re}[\sigma_{xy}(0)]=\frac{2}{\pi}\int_0^\infty
\frac{\mathrm{Im}[\sigma_{xy}(\omega)]}{\omega}d\omega .
\label{eq:KK}
\end{equation}

\begin{figure}[t!]
\centering
\includegraphics[width=0.5\textwidth]{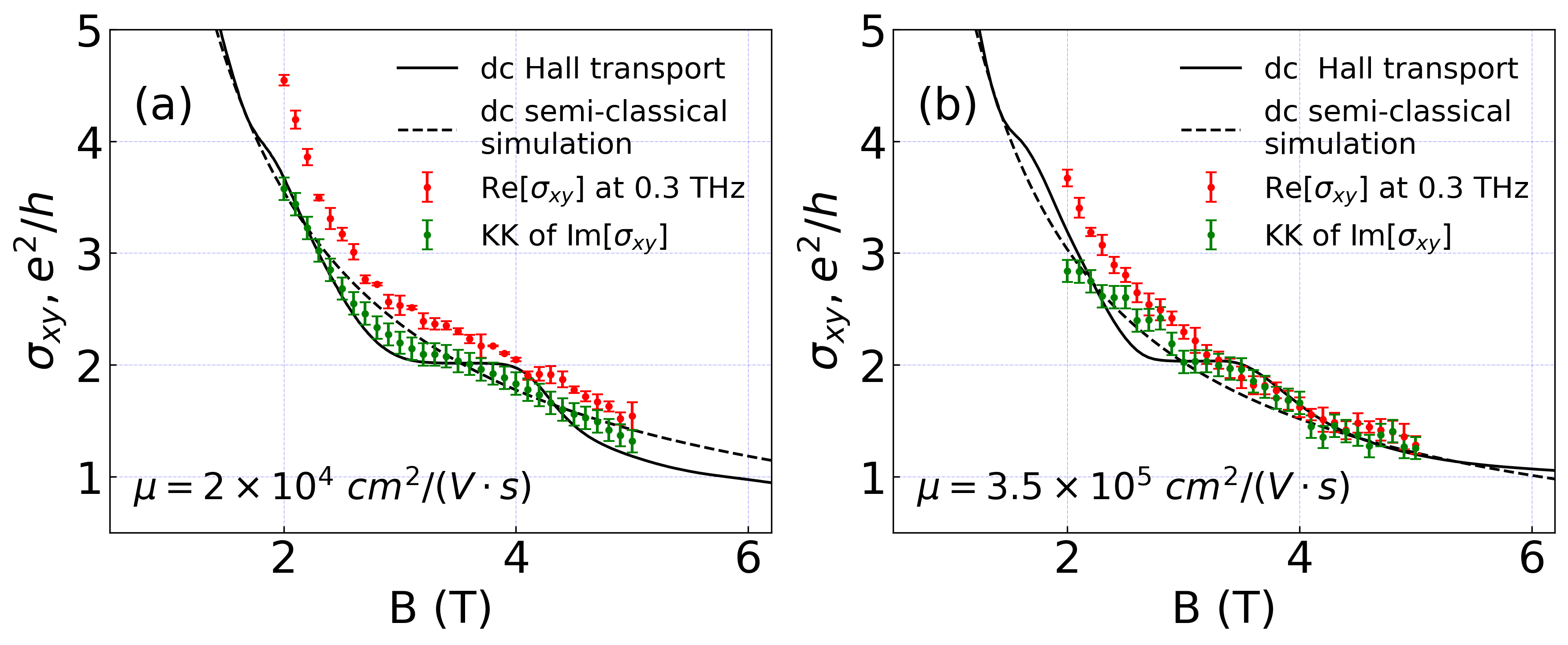} 
\label{Fig.Conductivity}
\caption{ {\bf Comparison between dc and THz Hall measurements as a function of magnetic field} for (a) lower-mobility and (b) higher-mobility 2DEG samples. Solid black lines indicate dc Hall transport measurements. Dashed black line shows semi-classical dc Hall prediction based on extracted carrier density. Data in red corresponds to Re[$\sigma_{xy}$] at lowest part of our spectral range 0.3 THz. Green dots represent Kramers-Kronig (KK) transformation of dissipative Im[$\sigma_{xy}$] calculated over 0.3~-~2.5~THz spectral range. }
\end{figure}
\begin{figure*}[t]
\centering
\includegraphics[width=0.8\textwidth]{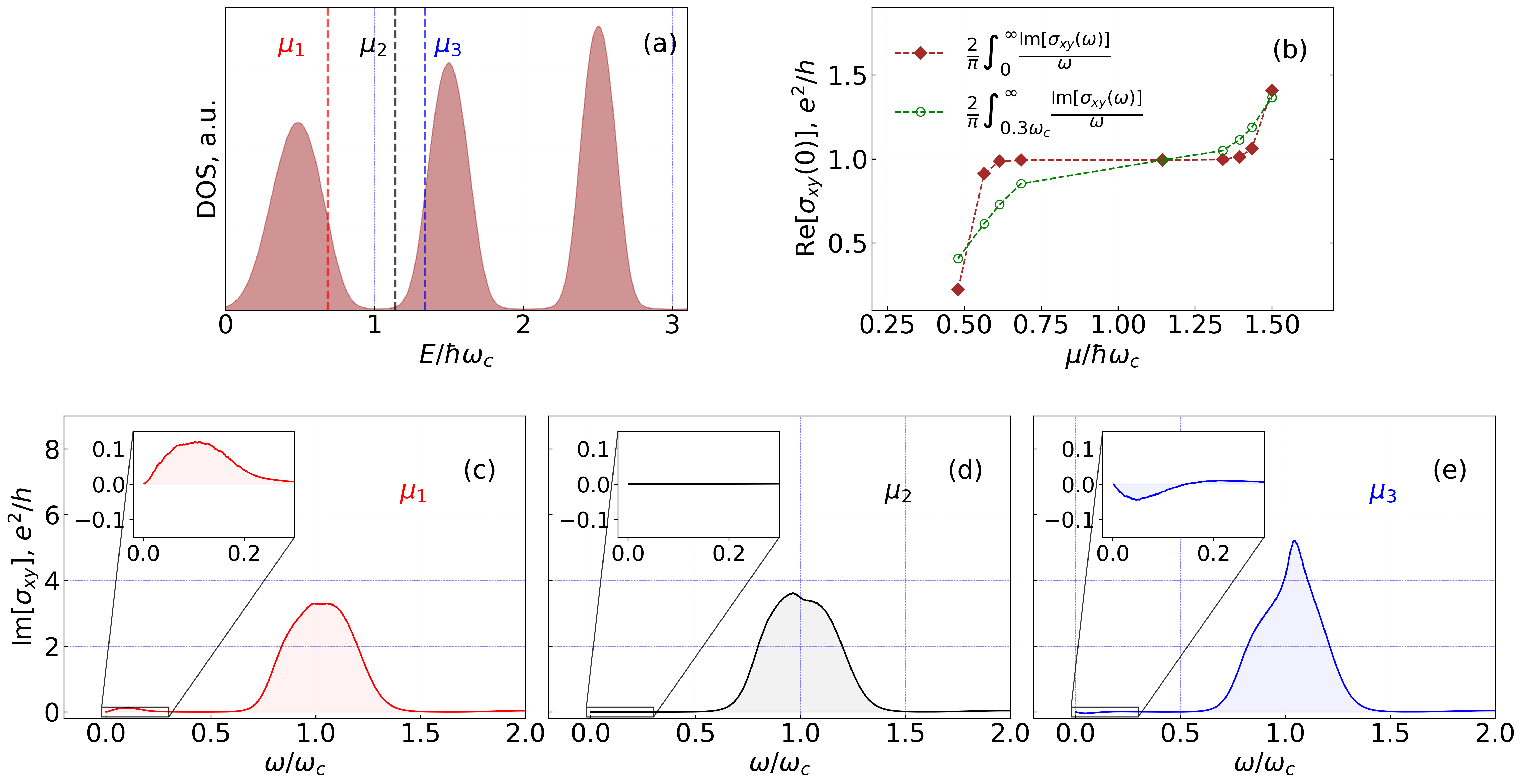} 
\label{Fig.Conductivity}
\caption{{\bf Numerical simulations} (a) Density of states $D(E)$. (b) Re[$\sigma_{xy}(0)$] computed from Kramers-Kronig of Im[$\sigma_{xy}(\omega)$] for various chemical potentials $\mu$. Red diamonds indicate the integration over entire $\omega$ range, green circles show integration which excludes low frequency range $\omega<0.3\omega_c$ (c)-(e) Im[$\sigma_{xy}(\omega)$] computed via Kubo formula for 3 different chemical potentials shown in (a). Insets show magnified views of the low-frequency contributions.}
\end{figure*}
In practice, the reconstruction does not require access to the full spectrum up to arbitrarily high frequencies, since contributions at high frequencies are suppressed by the $1/\omega$ factor in the  integral. The low-frequency part of the spectrum, on the other hand gives a much larger contribution. We use the experimentally accessible Im[$\sigma_{xy}(\omega)$] to reconstruct the zero-frequency response, with the resulting Re[$\sigma_{xy}$] shown by the green line in Fig.~3. Although it largely overlaps with the semi-classical $\sigma_{xy}^{\mathrm{class}}$, it does not reproduce the plateaus {\it actually} observed in DC transport. This demonstrates that although the high frequency cyclotron resonance dominates the $\sigma_{xy}(\omega)$ spectrum, the missing low frequency $0$--$0.3$ THz range must be crucial for recovering the DC quantum Hall response via a KK transform.  

Low-frequency contributions to Im[$\sigma_{xy}$] associated with disorder have in fact been observed previously with microwaves. Arakawa \textit{et al.} measured the dynamical Hall conductivity in the microwave regime and observed an oscillatory Im[$\sigma_{xy}$] that changes sign across the center of each Landau level \cite{Arakawa_Microwave}. They attributed this response to intra-Landau-level transitions between disorder-localized states with adjacent orbital angular momenta, $l_z\rightarrow l_z\pm1$ \cite{Yoshioka}. This demonstrates that disorder can generate a finite-frequency Hall response at energies far below the cyclotron resonance. Motivated by these observations, we investigated whether such impurity-induced low-frequency contributions can provide the missing spectral response required to recover the quantized DC Hall conductivity.  For this purpose, we perform numerical calculations for Landau levels subject to Gaussian-correlated disorder. We consider the disorder potential
\begin{equation}
V_{\mathrm{dis}}(\mathbf{r})=V_0\sum_{j=1}^{N_{\mathrm{imp}}}s_j
\exp\left[-\frac{|\mathbf{r}-\mathbf{R}_j|^2}{2\xi^2}\right],
\end{equation}
where the impurity positions $\mathbf{R}_j$ are chosen randomly and with $s_j=\pm 1$. The calculations are performed on a system containing 625 flux quanta, with 2000 positive and 2000 negative impurities. The impurity width and strength are fixed at $\xi=0.5\ell_B$ and $V_0=0.1\hbar\omega_c$, respectively, and the results are averaged over 1000 disorder realizations.

Fig.~4(a) shows the resulting calculated density of states $D(E)$ for the Landau levels (LLs) in the presence of disorder. The centers of the LLs remain at $\hbar\omega_c(\nu+\frac{1}{2})$, while their widths become narrower with increasing energy. In Figs.~4(c)-(e), we show the calculated frequency dependent Im[$\sigma_{xy}(\omega)$] for three different values of the chemical potentials: below the center of the $\nu=1$ quantum Hall plateau, near its center ($\mu=\hbar\omega_c$), and above it. In all cases, the ac response is dominated by cyclotron absorption, although its amplitude and width vary with the chemical potential.

Upon closer inspection, we note that when $\mu$ is away from the center of the $\nu=1$ plateau, Im[$\sigma_{xy}(\omega)$] develops an additional low-frequency contribution that is associated with impurity-induced transitions. One can see that while (for these simulations) the cyclotron resonance is always positive, the low frequency contribution is positive for $\mu$ below the center of the plateau and predominantly negative above it, while it is absent for $\mu\approx\hbar\omega_c$. 

To assess the significance of these transitions, we calculate Re[$\sigma_{xy}(0)$] in exactly the same way as we did for the experimental data i.e. using the KK relation of Eq.~\ref{eq:KK} for a range of chemical potentials. The results are shown by the brown diamonds in Fig.~4(b). As expected, the KK transformation recovers the quantized $e^2/h$ plateau for $\mu\approx0.6\hbar\omega_c-1.4\hbar\omega_c$ when integrating the full spectrum. To isolate the contribution of the impurity-induced low-frequency response, we repeat the calculation while excluding the $\omega<0.3\omega_c$ region from the KK integral. The resulting values are shown by the green circles in Fig.~4(b). It is clear that removing the low-frequency contribution leads to a significant deviation from the quantized plateau.   It is notable that for these simulations the cyclotron resonance range excitations clearly know about quantum Hall plateaus, but cannot by themselves be the entire contribution.   The impurity states are essential.

As can be seen in Figs.~4(c)-(e), these impurity-induced transitions carry extremely small spectral weight compared with the dominant cyclotron resonance. Nevertheless, because of their low characteristic frequencies and the corresponding $1/\omega$ weighting in the KK relation, their contribution to Re[$\sigma_{xy}(0)$] can reach as much as $0.3e^2/h$. Most remarkably, this contribution is not arbitrary. As the chemical potential moves across the Landau level and the cyclotron response changes, both the sign and spectral weight of the low-frequency contribution change by precisely the amount required to compensate for the non-quantized cyclotron response. Thus, these seemingly weak impurity-induced transitions provide precisely the missing contribution needed to recover the quantized $e^2/h$ Hall conductivity. Without this contribution, the Hall conductivity inferred from the finite-frequency response would not be quantized. 

In this work, we have shown that while the THz-range Hall conductance is dominated by a smoothly field-dependent cyclotron resonance that on its own gives only rounded, non-quantized features under Kramers-Kronig transformation, experiments and numerics demonstrate that a faint, low-frequency, topologically enforced dissipative contribution from impurity states supplies exactly the correction needed to recover the quantized plateau in the Kramers-Kronig integral.  This is corroborated by exact-diagonalization calculations of disordered Landau levels, which reproduce the quantized $e^2/h$ value only when this low-frequency response is retained in the Kramers-Kronig integral.  This gives a new route to understand quantum Hall quantization, one that is rooted in the dissipative finite-frequency electrodynamics of the disordered bulk.  Our results establish the direct, quantitative link between the finite-frequency optical response and the DC transport, and suggest that finite-frequency spectroscopy can serve as an independent experimental probe of the microscopic origin of quantization — one that is sensitive to disorder in a way the DC plateaus themselves are not. 

\bibliography{main} 
\section{Methods}
For DC transport measurements, electrical contacts were made using eutectic InSn. Small InSn contacts were placed at the four corners of a $4~\mathrm{mm}\times4~\mathrm{mm}$ sample and subsequently annealed at $425^\circ$C in a forming-gas ($\mathrm{H}_2$:$\mathrm{N}_2$) atmosphere. Transport measurements were performed in the van der Pauw geometry using an SR830 lock-in amplifier with an excitation current of $10~\mathrm{nA}$ at a frequency of $13~\mathrm{Hz}$. To remove contributions from imperfect contact alignment and geometric asymmetry, $R_{xy}$ was antisymmetrized with respect to the applied magnetic field, while $R_{xx}$ was symmetrized and corrected by the appropriate geometric factor. The DC Hall conductance was obtained from the measured longitudinal and Hall resistances according to
\begin{equation}
    \sigma_{xy}=\frac{R_{xy}}{R_{xx}^2+R_{xy}^2}.
\end{equation}
A full DC characterization of both samples is presented in Supplementary Fig.~S1.

The finite-frequency Hall conductance $\sigma_{xy}(\omega)$ was measured using high-precision time-domain terahertz polarimetry (TDTP). Experimental setup details can be found in Ref.~\cite{HighPrecision}. By performing TDTP measurements on the sample and a bare substrate, we obtain all elements of the sample transfer matrix $\hat{T}$. Since TDTP is a phase-sensitive probe, the complex tensorial response is obtained directly from the experimental data. In the presence of a magnetic field, the natural eigenbasis for the interaction of electromagnetic radiation with the 2DEG is the circular polarization basis. Although our measurements are performed in the Cartesian basis, we define the effective transmission for right ($r$) and left ($l$) circular polarizations as $T_{rr/ll}=T_{xx}\pm iT_{xy}$. The corresponding conductivities are then obtained using the usual thin-film relation for a conducting film on a dielectric substrate,

\begin{equation}
\sigma_{rr/ll}(\omega)=\frac{n+1}{Z_0}
\left(
\frac{e^{i\Delta\phi}}{T_{rr/ll}(\omega)}-1
\right),
\end{equation}
where $n$ is the refractive index of the substrate, $Z_0$ is the impedance of free space, and $\Delta\phi$ accounts for the phase difference between the sample and reference substrates. In the context of QH systems, it is more convenient to work with the longitudinal ($\sigma_{xx}$) and transverse ($\sigma_{xy}$) conductances rather than their circular-basis counterparts, as we have done in Refs.~\cite{THz_QAHE, AnaelleLSCO, BingCdAsPhonon}. We therefore convert the extracted circular conductivities back to the Cartesian basis using $\sigma_{xx}=(\sigma_{rr}+\sigma_{ll})/2$ and $\sigma_{xy}=(\sigma_{rr}-\sigma_{ll})/2i$. Thus, the circular basis is used only as an intermediate step in extracting the conductivity.  Throughout this work we present and discuss the response in terms of $\sigma_{xx}$ and $\sigma_{xy}$.  By defining the complex Faraday rotation as $\hat{\theta}=T_{xy}/T_{xx}$ and making the practical assumption that $T_{xy}\ll T_{xx}$, one obtains

\begin{equation}
\sigma_{xy}(\omega)=\frac{n+1}{Z_0 }\frac{\hat{\theta}(\omega)}{T_{xx}(\omega)}.
\end{equation}

 \section{Acknowledgments: }
 
The work at JHU was supported by the Army Research Office MURI ``Implementation of axion electrodynamics in topological films and device" W911NF2020166.  Instrumentation development at JHU, which made these measurements possible was supported by the Gordon and Betty Moore Foundation EPiQS Initiative Grant GBMF-9454 to NPA. The Princeton portion of this work is funded in part by the Gordon and Betty Moore Foundation EPiQS initiative, Grant GBMF14260.  The work at Massachusetts Institute of Technology was supported by the Air Force Office of Scientific Research under award number FA2386-24-1-4043.  LF and NPA had additional support from the Quantum Materials program at the Canadian Institute for Advanced Research.
 
 \section{Author contributions:  }
 ZT performed the experiments and the analysis of the data.  AA and YO performed the numerical simulations and theory analysis.  AG and LP grew the samples.  LF and NPA directed the project.   All authors contributed to the writing and editing of the manuscript.

 \section{Additional information:} 
 
\textbf{Competing interests:} The authors declare no competing interests.

\clearpage
\onecolumngrid
\setcounter{figure}{0}
\renewcommand{\thefigure}{S\arabic{figure}}

\title{\LARGE{Supplementary Figures} 
\\
{\LARGE \textbf{Quantization through dissipation: impurity states and the optical Hall effect}} }

%\author[1]{Zhenisbek Tagay}
%\author[2]{Ahmed Abouelkomsan}
%\author[2]{Yugo Onishi}
%\author[3]{Adbhut Gupta}
%\author[3]{Loren Pfeiffer}
%\author[2]{Liang Fu}
%\author[1]{N.~P. Armitage}
%\affil[1]{\textit{Department of Physics and Astronomy, The Johns Hopkins University, Baltimore, MD 21218 USA}}
%\affil[2]{\textit{Department of Physics, Massachusetts Institute of Technology, Cambridge, Massachusetts 02139, USA}}
%\affil[3]{\textit{Department of Electrical and Computer Engineering, Princeton University, Princeton, NJ 08854 USA}}

%\date{\vspace{-5ex}}

%\title{Supplemental Material}
%\maketitle
\begin{center}
    {\LARGE \textbf{Supplementary Figures}}
\end{center}
\begin{figure}[h!]
\centering
\includegraphics[width=0.7\textwidth]{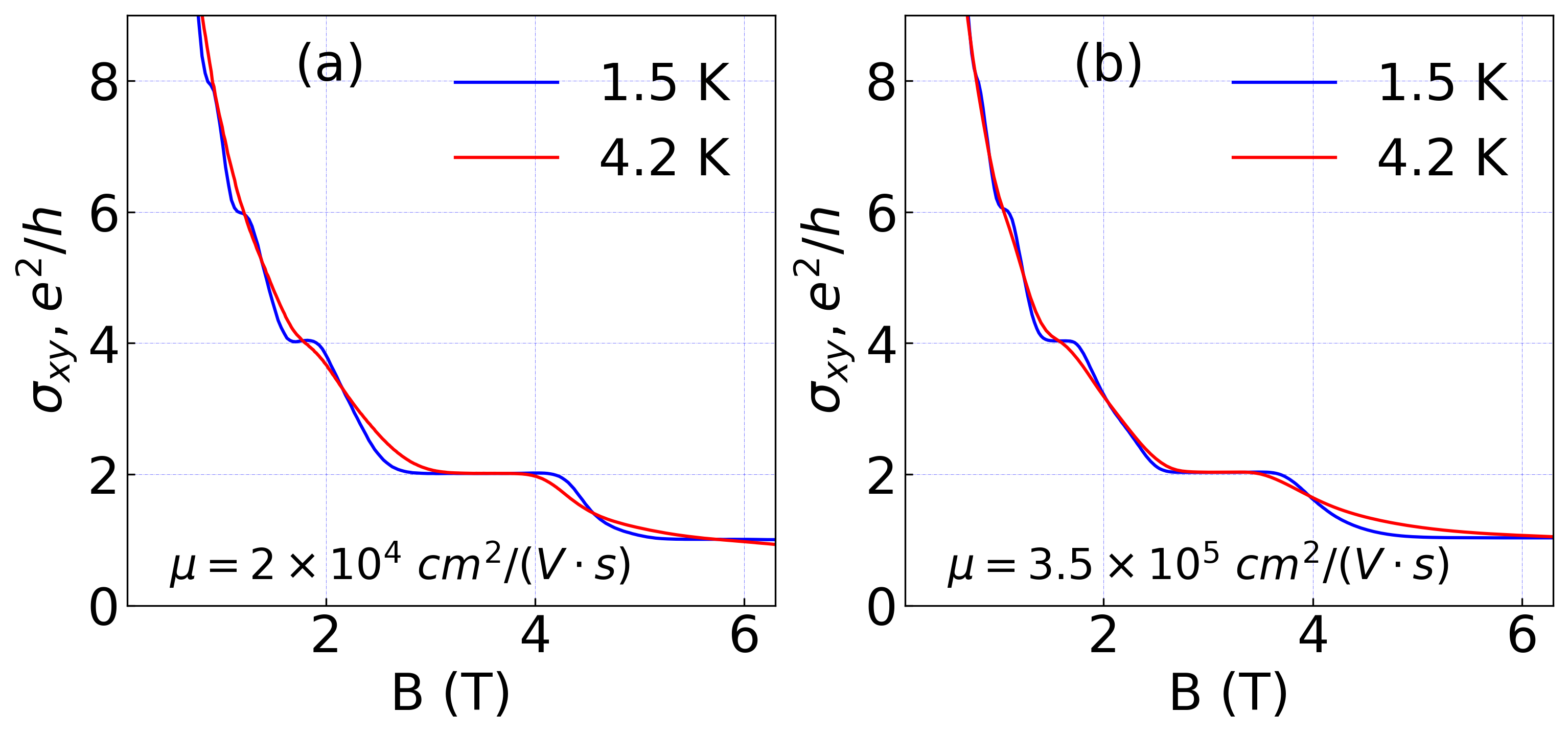} 
\label{Fig.Conductivity}
\caption{{\bf DC Hall conductance} for lower- (a) and higher-mobility (b) samples, respectively. Both the $1.5~\mathrm{K}$ and $4.2~\mathrm{K}$ data exhibit a well-defined $\nu=2$ plateau, with the plateau extending over a slightly wider magnetic-field range at the lower temperature. At $1.5~\mathrm{K}$, additional plateaus corresponding to $\nu=1,4,6,$ and $8$ are also resolved.}
\end{figure}

\begin{figure}[h!]
\centering
\includegraphics[width=0.7\textwidth]{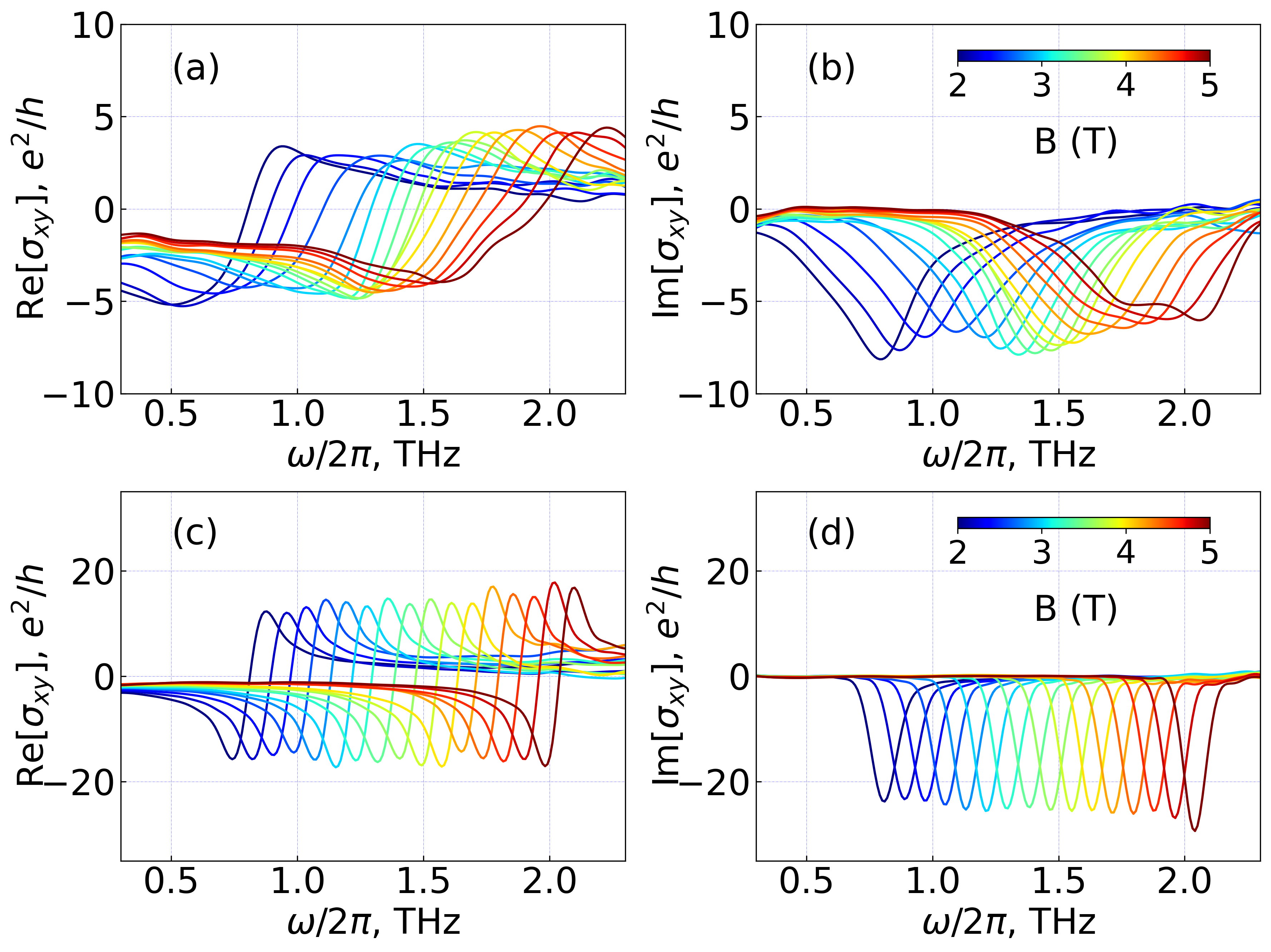} 
\label{Fig.Conductivity}
\caption{{\bf THz range complex Hall conductivity at T~=~1.5~K} (a) Real and (b) imaginary Hall conductance $\sigma_{xy}(\omega)$ of lower-mobility 2DEG at various magnetic fields.(c)-(d) Same thing for higher-mobility 2DEG.}
\end{figure}

\begin{figure}[h]
\centering
\includegraphics[width=0.9\textwidth]{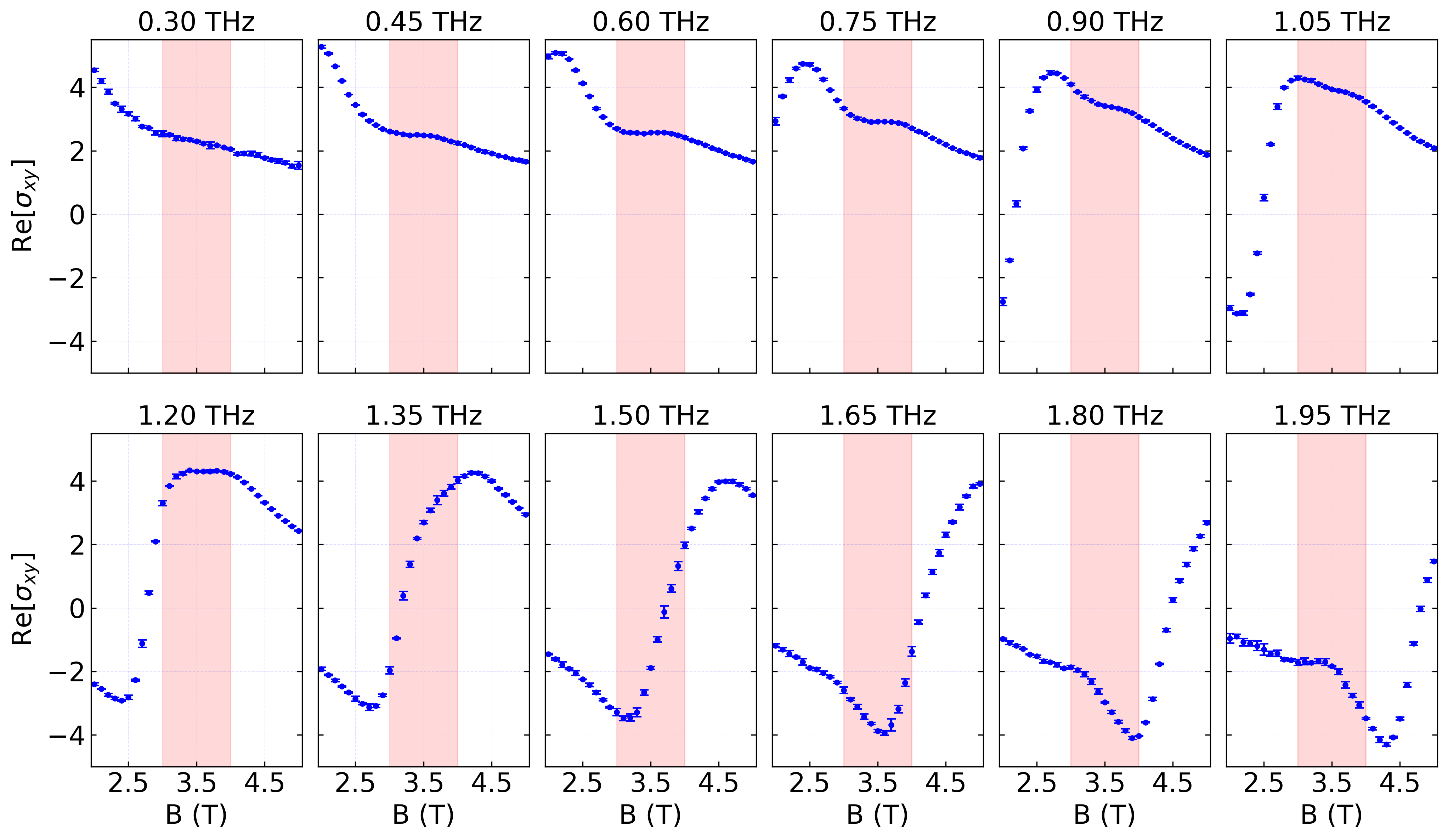} 
\label{Fig.Conductivity}
\caption{{\bf Frequency cuts of Re[$\sigma_{xy}$] for lower-mobility sample at $T=4.2$ K.}}
\end{figure}

\begin{figure}[h]
\centering
\includegraphics[width=0.9\textwidth]{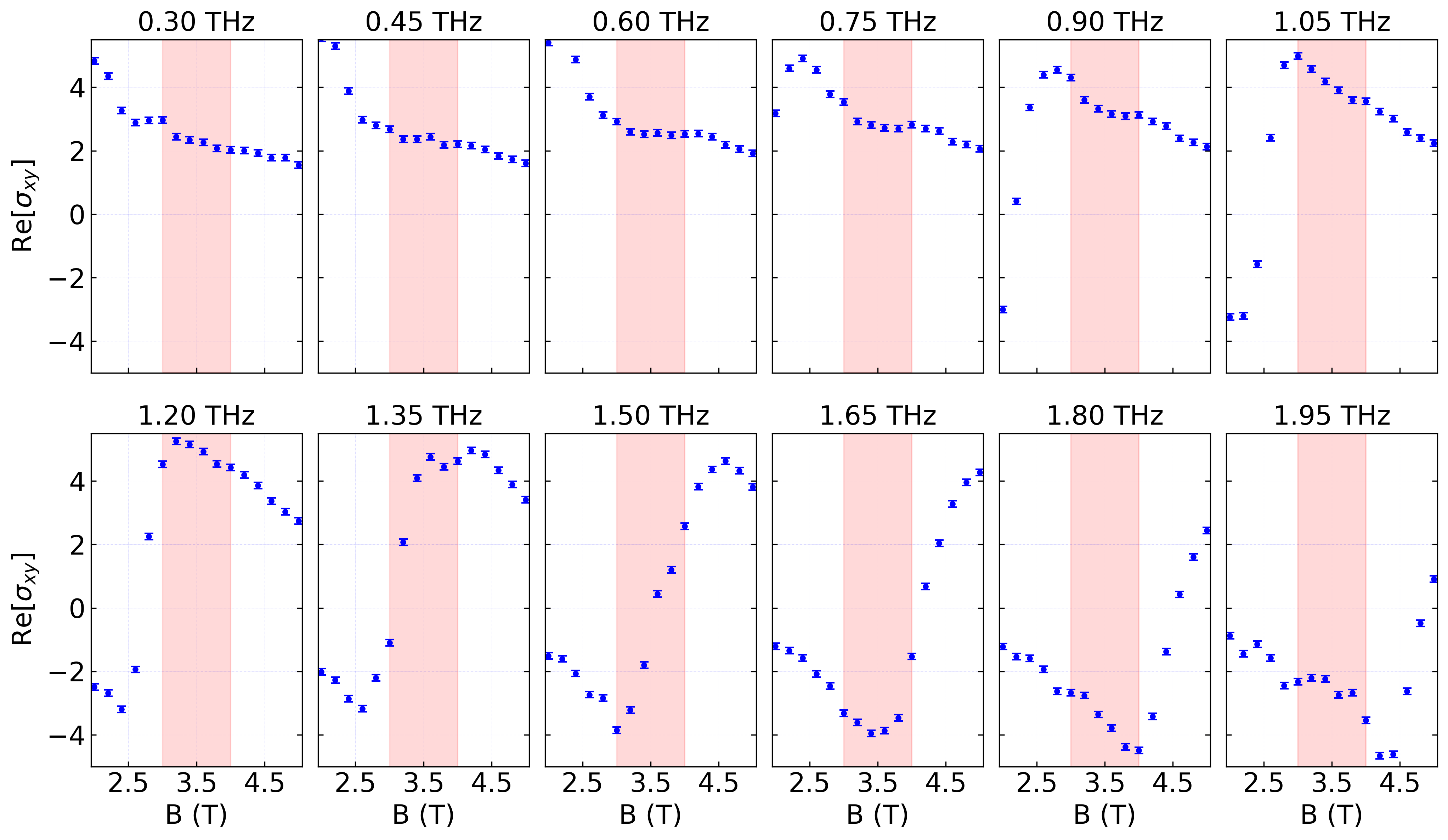} 
\label{Fig.Conductivity}
\caption{{\bf Frequency cuts of Re[$\sigma_{xy}$] for lower-mobility sample at $T=1.5$ K.}}
\end{figure}

\begin{figure}[h]
\centering
\includegraphics[width=0.9\textwidth]{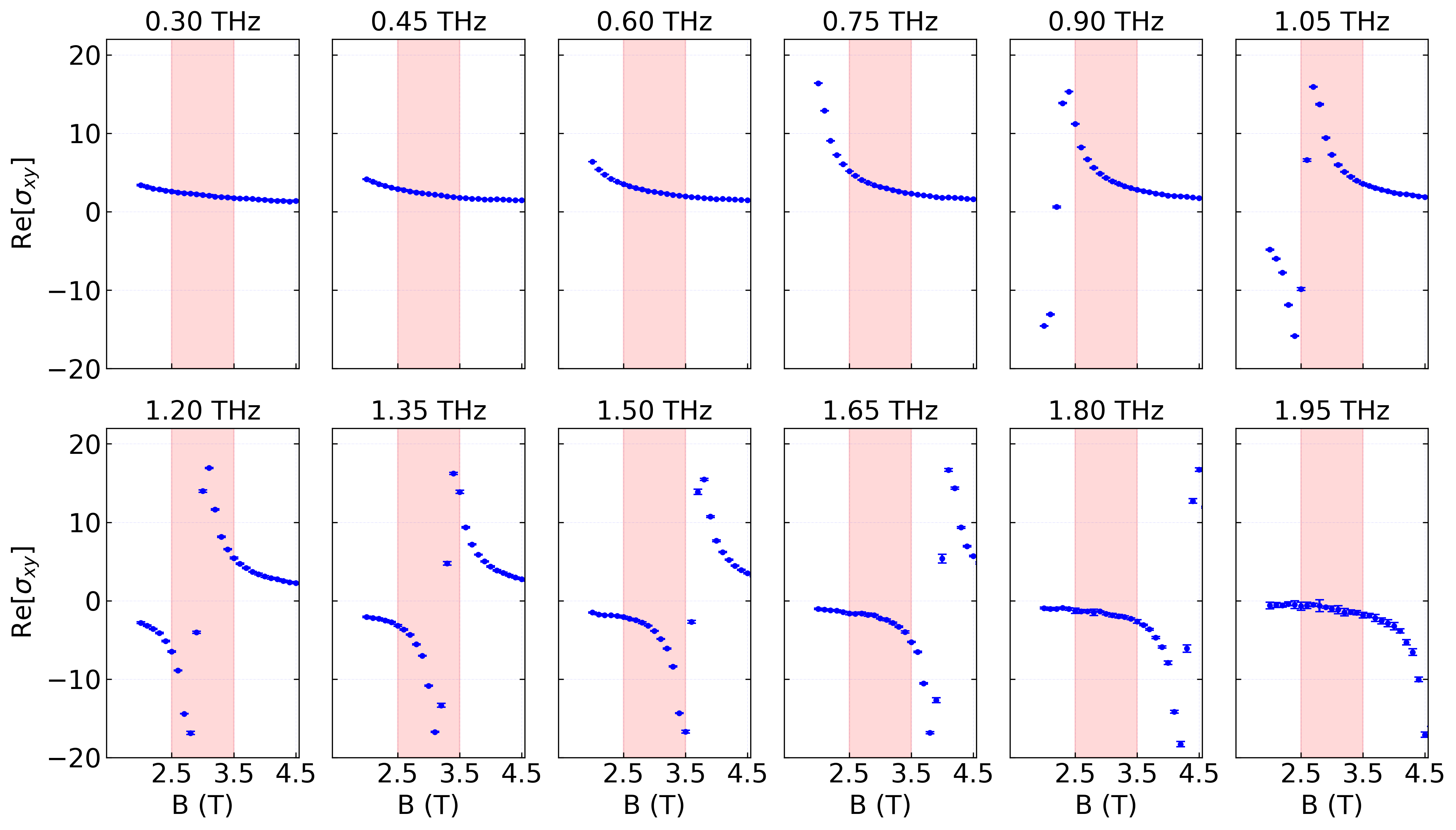} 
\label{Fig.Conductivity}
\caption{{\bf Frequency cuts of Re[$\sigma_{xy}$] for higher-mobility sample at $T=4.2$ K.}}
\end{figure}

\begin{figure}[h]
\centering
\includegraphics[width=0.9\textwidth]{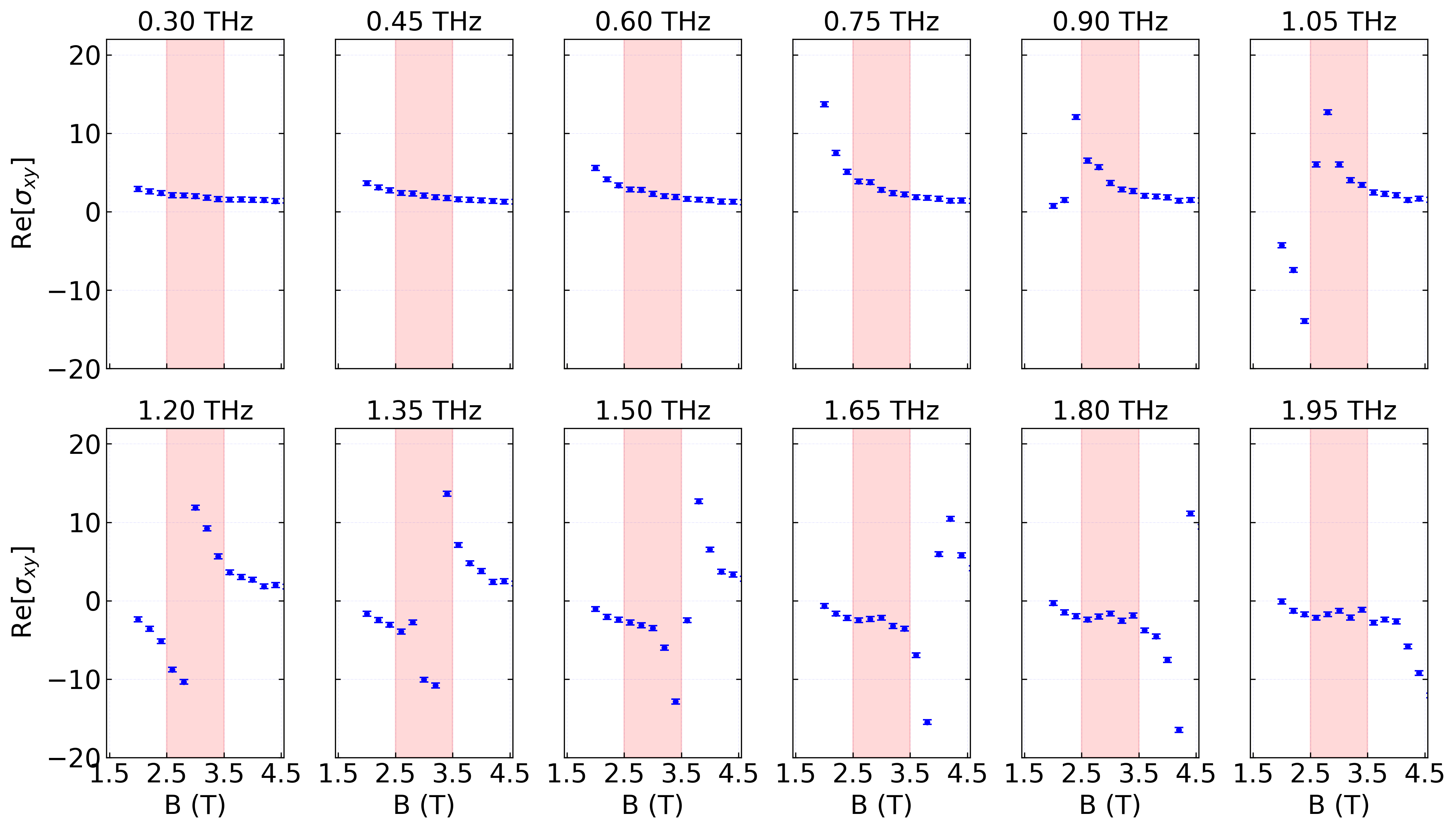} 
\label{Fig.Conductivity}
\caption{{\bf Frequency cuts of Re[$\sigma_{xy}$] for higher-mobility sample at $T=1.5$ K.}}
\end{figure}

\end{document}